\documentclass[traditabstract]{aa}
\usepackage{graphicx}
\usepackage{natbib}
\usepackage{txfonts}
\bibpunct{(}{)}{;}{a}{}{,}
\begin{document}

  \title{Post first dredge-up [C/N] ratio as age indicator. Theoretical calibration}
  
  \author{Maurizio Salaris\inst{1}, Adriano Pietrinferni\inst{2}, Anna M. Piersimoni\inst{2} \and Santi Cassisi\inst{2,3}}

\institute{Astrophysics Research Institute, 
           Liverpool John Moores University, 
           IC2, Liverpool Science Park, 
           146 Brownlow Hill, 
           Liverpool L3 5RF, UK, \email{M.Salaris@ljmu.ac.uk} 
            \and INAF~$-$~Osservatorio Astronomico di Collurania, Via M. Maggini, I$-$64100 , Teramo, Italy, 
            \email{pietrinferni,piersimoni,cassisi@oa-teramo.inaf.it}  
            \and   Instituto de Astrof{\'i}sica de Canarias, Calle Via Lactea s/n, E-38205 La Laguna, Tenerife, Spain
           }

 \abstract{
We performed a detailed analysis of the use of [C/N] measured in red giant branch stars between 
the completion of the first dredge up and the red giant branch bump (${\rm [C/N]_{FDU}}$) as age indicator. 
${\rm [C/N]_{FDU}}$ cannot give accurate ages for individual stars, but may  
provide a general chronology for the formation of composite populations and add constraints to analyses of red giants    
from surface gravity-effective temperature diagrams.  
We provide a theoretical calibration of ${\rm [C/N]_{FDU}}$ in terms of total metallicity [M/H] and age,
for ages greater than $\sim$1~Gyr, which we tested against variations in the initial heavy element distribution (scaled-solar 
vs $\alpha$-enhanced), efficiency of overshooting from MS convective cores and from the convective envelopes, variations 
in the initial He abundance and in the mixing length parameter.
Our calibration is compared with a small sample of available measurements of ${\rm [C/N]_{FDU}}$ 
in star clusters and halo field stars, which at least qualitatively confirm the overall trend of 
the predicted ${\rm [C/N]_{FDU}}$ with age and [M/H]. 
The use of ${\rm [C/N]_{FDU}}$-[M/H]-age relations obtained from independent sets of stellar evolution calculations 
cause age differences (for a given ${\rm [C/N]_{FDU}}$ and [M/H] pair) up to $\sim$2~Gyr. 
More accurate spectroscopic measurements of ${\rm [C/N]_{FDU}}$ in star clusters with well-established 
ages and metallicities are required to better test theoretical calibrations of this age indicator.
}
\authorrunning{M. Salaris et al.}
\titlerunning{Post first dredge-up [C/N] ratio as age indicator}
  \maketitle


\section{Introduction}
Large spectroscopic surveys, such as
LAMOST \citep[Large sky Area Multi-Object fiber Spectroscopic Telescope,][]{lamost}, 
SEGUE \citep[Sloan Extension for Galactic Understanding and Exploration,][]{segue}, 
GAIA-ESO \citep[][]{gaiaeso}, and 
APOGEE \citep[Apache Point Observatory Galactic Evolution Experiment,][]{apogee}  
have been devised to provide surface gravity, effective temperature,
radial velocities, and photospheric abundances of several chemical
elements for large samples of Milky Way stars, with the aim of 
unravelling our galaxy detailed evolutionary history. 
In parallel, the $Kepler$ mission \citep{keplera, keplerb} provides asteroseismic data for dwarfs and red giants, 
which lead to an  estimation of stellar masses and radii. 

Translating this data into an evolutionary history of the Galaxy requires  
age estimates of the observed field stars. These are usually performed by fitting
theoretical stellar evolution tracks to the observations 
in the log($g$)-log(${T_{\mathrm eff}}$) diagram, with additional constraints given by
the measured chemical abundances and asteroseismic measurements when available  
\citep[see, e.g.,][]{s13, r14, v15}.
Ages of red giant branch (RGB) stars are particularly challenging because small uncertainties
in $T_{\mathrm eff}$ (at fixed $g$) translate into large uncertainties on the star's mass, hence
its age, especially when asteroseismic constraints on the mass of the object are not available.
Additional independent constraints on the age of RGB stars 
are therefore welcome, to be used in conjunction with the log($g$)-log(${T_{\mathrm eff}}$) diagram
(and asteroseismic masses) to improve the accuracy of the age estimates.

\cite{mt15} have recently constrained the age ranges spanned by thin and thick disk RGB stars 
using the [C/N] ratio\footnote{As is customary, we follow the spectroscopic notation 
for the ratio [A/B] 
between the number fractions of elements A and B, given as  
[A/B]=${\rm log}(n_{\mathrm A}/n_{\mathrm B})_{star}-{\rm log}(n_{\mathrm A}/n_{\mathrm B})_{\odot}$. 
The value of [A/B] is the same if element mass fractions (usually employed in 
stellar evolution calculations) are used instead of number fractions (typically 
determined from spectroscopic observations).} 
measured in RGB stars (from the APOGEE survey) 
after the completion of the first dredge-up (FDU), but fainter than the RGB bump luminosity, 
where observations \citep[see, e.g.,][]{g00, a15} reveal the onset of an extra mixing process which continues until the tip of
the RGB (and possibly beyond). Theoretical stellar models predict that this ratio --hereafter ${\rm [C/N]_{FDU}}$-- depends on the mass of the star
(hence its age) and to some degree on its initial chemical composition.
We present here 
a theoretical analysis of its cali\-bra\-tion in terms of total metallicity [M/H] and age t,
for t more than $\sim$1~Gyr, e.g.  
for stellar populations with well-populated RGBs, which host stars with an electron
degenerate He-core (denoted here as low-mass stars).

The next section describes the evolution of the [C/N] ratio during the FDU, provides
a theoretical calibration of the ${\rm [C/N]_{FDU}}$-[M/H]-t relationship for low-mass stars which is compared 
with observations, and discusses several theoretical uncertainties relevant to the calibration. 
Section~3 compares our results with ${\rm [C/N]_{FDU}}$-[M/H]-t relationships determined from
independent sets of stellar models, and this is followed by a summary of our results.

\section{Theoretical calibration of ${\rm [C/N]_{FDU}}$ vs age}
 
Stellar model calculations show that at the end of the main sequence (MS) phase, the   
outer convection zone progressively engulfs deeper regions, dredging  
to the surface matter which has been partially processed by H-burning during the MS (the FDU). 
Chemical elements affected by the FDU are essentially C, N, Li, and He. Both He and N abundances  
increase, whilst Li and C decrease, the ${\rm ^{12}C/ ^{13}C}$ ratio drops to values of $\sim$20-25 
\citep[see, e.g.,][and references therein]{kl} and the ${\rm^{14}N/ ^{15}N}$ ratio increases, 
due to mixing of the envelope's initial composition with inner layers processed by the CN-cycle, where Li has also been burned.
This slow variation in the photospheric chemical composition 
halts at the maximum penetration of the convective envelope, at which point the inner convective boundary 
starts to move back towards the surface, leaving behind a chemical discontinuity. This discontinuity, when 
crossed by the advancing (in mass) H-burning shell, causes the appearance of the RGB bump in the luminosity function 
of old stellar populations \citep[see, e.g.][]{cs13}.

\begin{figure}
\centering
\includegraphics[scale=.4500]{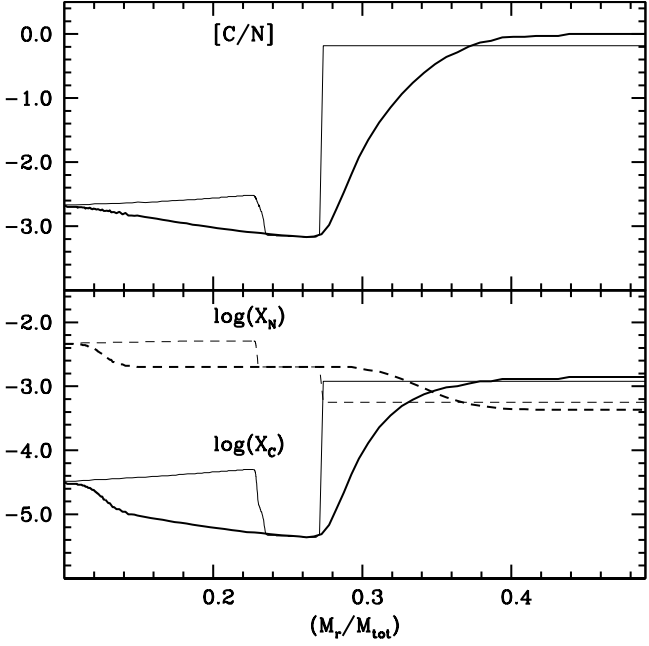}
\caption{{\sl Upper panel}: run of the [C/N] abundance ratio within 
  a 0.9${\rm M_{\odot}}$ stellar model ([M/H]=$-$0.35, scaled solar metal mixture), before the occurrence of the FDU (thick line)
  and right after the completion of the FDU (thin line). The horizontal axis displays the local fractional mass
  (mass enclosed within a distance $r$ from the centre divided by the total mass). 
  {\sl Lower panel}: as upper panel but for the individual mass fractions of
  N (dashed lines) and C (solid lines).}
\label{int_a}
\end{figure}

As an example, Fig.~\ref{int_a} displays the inner C and N stratifications together
with the corresponding [C/N] ratio before the start, 
and at the completion of the FDU respectively, of a 0.9$M_{\odot}$ model with metallicity [M/H]=$-$0.35 
from the BaSTI stellar evolution database\footnote{http://www.oa-teramo.inaf.it/BASTI/} 
\citep{basti, bastialpha}
This value of [M/H]  
is obtained from $Z$=0.008, $Y$=0.256 (scaled solar metal mixture) 
according to [M/H]=${\rm log}(Z/X)_{star}-{\rm log}(Z/X)_{\odot}$ where $Z$, $X=1-Z-Y$, and $Y$ denote the 
metal, hydrogen and He mass fractions\footnote{In general, if the metal mixture of the models is scaled solar, [M/H]=[Fe/H]. In case of 
$\alpha$-element enhanced ([$\alpha$/Fe]$>$0) metal distributions, to a good approximation 
${\rm [M/H]\sim [Fe/H]+log(0.638 f_{\alpha}+0.362)}$ where $f_{\alpha}=10^{\rm [\alpha/Fe]}$ \citep[][]{scs}.}   

The pre-FDU nitrogen chemical profile displays clearly three steps. Moving 
inwards, we find the initial abundance first, then the higher CN-cycle equilibrium value, and the even higher full
CNO-cycle equilibrium value in the very central layers. 
The carbon abundances follow their own pattern of initial, CN-cycle and CNO-cycle equilibrium abundances. 
After the FDU is completed, convection has mixed the surface with the CN equilibrium layers, and the final uniform abundance   
of N in the envelope has increased, whilst the C abundance has decreased.
The corresponding behaviour of the [C/N] ratio is displayed in the upper panel of the same figure.

At FDU completion the convective envelope   
contains $\sim$72\% of the total mass. When the mass of the model increases, 
surface convection at the FDU engulfs a larger fraction of the total stellar mass, 
and the net result is 
a decrease in [C/N] with increasing RGB mass --hence decreasing age of the host population-- 
in the relevant mass (age) range.

\begin{figure}
\centering
\includegraphics[scale=.4500]{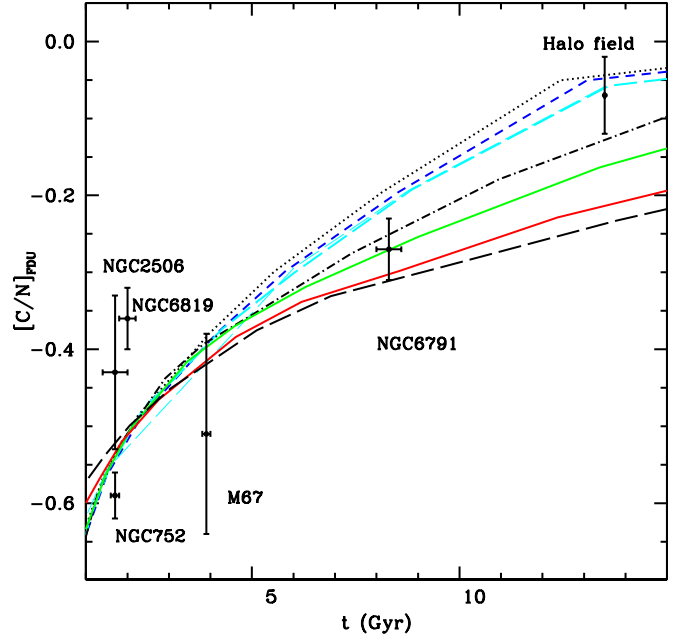}
\caption{Theoretical predictions for the run of ${\rm [C/N]_{FDU}}$ as a function of age for selected metallicities.
  From top to bottom, at an age of 10~Gyr, [M/H] is equal to
  $-$2.27 (dotted line), $-$1.49 (short dashed), $-$1.27(long dashed), $-$0.66
  (dot-dashed),  $-$0.35 (thick solid), 0.06 (thin solid) and 0.26~dex (thin long dashed), respectively.
  Data for halo field stars and the labelled open clusters are also plotted
  (see text for details). }
\label{calibration}
\end{figure}

Figure~\ref{calibration} displays our theoretical calibration of the ${\rm [C/N]_{FDU}}$-[M/H]-t relation,
obtained from the BaSTI stellar model library, including overshooting from MS convective cores 
and a scaled-solar heavy element distribution \citep[see][ for more details]{basti}. 
The MS convective cores have been extended beyond the 
boundary given by the Schwarzschild criterion by 
an amount $\lambda_{ov} H_p$, where $H_p$ is the pressure scale height at the Schwarzschild border and 
$\lambda_{ov}$ a free parameter.  
For masses larger than or equal to 1.7$M_{\odot}$, BaSTI models employ $\lambda_{ov}$=0.20$H_p$; for stars less massive than 
1.1$M_{\odot}$ $\lambda_{ov}$=0, while in the intermediate range 
of the model grid (M=1.1, 1.2, 1.3, 1.4, 1.5 and 1.6$M_{\odot}$, respectively) $\lambda_{ov}$
varies as $\lambda_{ov}$=($M/M_{\odot}$-0.9)/4 \citep[see][for details]{basti}. 

To determine the 
${\rm [C/N]_{FDU}}$-[M/H]-t calibration we have considered all low-mass model calculations at a given [M/H], 
and taken age and surface C and N abundances predicted for each mass at the RGB bump. The solar (C/N) ratio comes from the 
\citet{gn93} solar metal mixture employed in BaSTI calculations.
Notice that the age range between FDU completion and RGB bump is negligible. 

For t above $\sim$4~Gyr there is a clear dependence of ${\rm [C/N]_{FDU}}$ on [M/H], that increases
with incresing age.
Lower metallicities imply higher ${\rm [C/N]_{FDU}}$, due to the shallower convective envelopes 
and a consequent smaller overlap with the CN-equilibrium region.
At any given metallicity the dependence of ${\rm [C/N]_{FDU}}$ on t is highly non linear, the slope
$\Delta {\rm [C/N]_{FDU}}$/$\Delta$t increasing sharply for ages below $\sim$5~Gyr; 
at any age this slope is shallower with increasing metallicity.
As a consequence, the value of ${\rm [C/N]_{FDU}}$ becomes increasingly less affected by [M/H] when t decreases,
and it is predicted to be almost independent of [M/H] when t is below $\sim$3~Gyr.

${\rm [C/N]_{FDU}}$ measurements alone cannot however provide very precise ages of individual 
field stars, when considering typical errors of $\sim$0.10-0.15~dex \citep[see, e.g.,][]{n6791, n2506}.  
At [M/H]=$-$2.27 ${\rm [C/N]_{FDU}}$ varies by $\sim$0.60~dex for a total age range equal to $\sim$13~Gyr, whilst at 
[M/H]= 0.26 the  ${\rm [C/N]_{FDU}}$ variation over the same age range is just $\sim$0.35~dex. 
On the other hand, ${\rm [C/N]_{FDU}}$ can provide additional age constraints on individual RGB 
stars to be added to information coming from the
log($g$)-log(${T_{\mathrm eff}}$) diagram (and eventually masses determined from asteroseismology), 
and also a general chronology for the formation of composite populations, when large samples  
of ${\rm [C/N]_{FDU}}$ measurements are available, as done by \citet{mt15}.  
We have also considered the ${\rm [C/N]_{FDU}}$ values predicted by the BaSTI 
$\alpha-$enhanced stellar models \citep{bastialpha}, and found a  
${\rm [C/N]_{FDU}}$-[M/H]-t relation almost identical to the scaled solar one.

The following analytical fit (our reference calibration) provides the age of a star with measured ${\rm [C/N]_{FDU}}$ and [M/H], as obtained
from the BaSTI models discussed above:

$$ {\rm t (Gyr)= 34.99 + 156.3 [C/N]_{FDU} + 20.52 [M/H] + } $$
$$ {\rm 298.6 ([C/N]^2_{FDU}) + 78.16 [C/N]_{FDU} [M/H] + 7.82 ([M/H]^2) }$$
$${\rm  + 305.9 ([C/N]^3_{FDU}) + 85.02 ([C/N]^2_{FDU}) [M/H] }$$
$${\rm  + 22.93 [C/N]_{FDU} ([M/H]^2) + 0.8987 ([M/H]^3) }$$
$${\rm  + 141.1 ([C/N]^4_{FDU}) + 21.96 ([C/N]^3_{FDU}) [M/H] }$$
$${\rm  + 16.14 ([C/N]^2_{FDU}) ([M/H]^2) + 1.447 [C/N]_{FDU} ([M/H]^3) }$$

with an accuracy within $\pm$15\%, for ages between 0.9 and 14.5~Gyr, and the 
[M/H] range of Fig.~\ref{calibration}. It can be applied to 
stars fainter than the RGB bump, which have completed the FDU.  
The values of surface gravity as a function of ${\rm T_{\mathrm eff}}$ 
at the completion of the FDU ($g_{\rm FDU}$) and the RGB bump ($g_{\rm bump}$) predicted by the BaSTI models are given by:
$ {\rm log}(g_{{\rm FDU}})=-11.95 {\rm log}(T_{\mathrm eff})+47.10 $ and ${\rm log}(g_{{\rm bump}})=-14.44 {\rm log}(T_{\mathrm eff})+55.30 $, 
with $\sim$0.15~dex spread, for the [M/H] range of Fig~\ref{calibration}. 
Observed RGB stars that in a $g$-${\rm T_{\mathrm eff}}$ diagram lie between 
these relationships for $g_{\rm FDU}$ and $g_{\rm bump}$, can 
therefore be age-dated with our ${\rm [C/N]_{FDU}}$-[M/H]-t calibration.

The models employed for our reference ${\rm [C/N]_{FDU}}$-[M/H]-t relation do not include the effect of atomic diffusion 
and radiative levitation at old ages. Observations suggest that these processes are largely inhibited 
--at least from the convective envelopes-- 
in low mass metal poor stars \citep[see, e.g.,][and references therein]{korn, msl11}. Moreover, after the FDU the surface C/N ratio 
(and to a good approximation all chemical abundances) is basically the same as for the case of neglecting diffusion in low mass 
metal poor models, as shown for example by \citet{mrr}.
The effect of fully efficient diffusion 
would be just a decrease by about 5\% of the estimated ages at fixed ${\rm [C/N]_{FDU}}$ and [M/H].

Our reference calibration does not account for overshooting from the lower boundary of convective envelopes.
To check the impact of this process on ${\rm [C/N]_{FDU}}$, 
we have computed selected additional models at [M/H]=0.06 ($Z$=0.02, $Y$=0.273) extending the fully mixed envelope by
a distance $\lambda_{ov}=0.3H_P$, a choice consistent with indications from the discrepancy between
predicted and observed brightness of the RGB bump in Galactic globular clusters
\citep[see, e.g.][]{cassisi:11}.
We found that the change (decrease) of  
${\rm [C/N]_{FDU}}$ is generally very small, equal at most to $\sim$0.02~dex. 

We performed three additional tests at [M/H]=0.06. The first one is an assessment of the effect of neglecting
convective core overshooting during the MS. Small differences in ${\rm [C/N]_{FDU}}$ at a given t appear for t below $\sim$3~Gyr, 
and reach at most 0.05~dex (${\rm [C/N]_{FDU}}$ higher in models without overshooting) at t=1~Gyr. 
We have then calculated models to study the effect of varying $Y$ (a decrease ${\rm \Delta{Y}=0.02}$) and 
the mixing length parameter $\alpha_{\rm ml}$ (a decrease in $\alpha_{\rm ml}$ by 0.3 compared to the solar calibrated reference value
$\alpha_{\rm ml, \odot}$=2.01 of the BaSTI calculations) respectively. In both cases the predicted ${\rm [C/N]_{FDU}}$-[M/H]-t relation is unchanged.

We have then compared our reference theoretical calibration with observations of ${\rm [C/N]_{FDU}}$ 
in field halo stars and a few open clusters of different ages. 
Observational data are scarce and error bars sizable when compared to the scaling 
of ${\rm [C/N]_{FDU}}$ with age and metallicity, and do not allow a stringent test of the models.

To sample old and metal-poor systems we considered the field halo stars observed by \citet{g00}, with an estimated mean 
${\rm [C/N]_{FDU}}=-0.07\pm0.05$  (error equal to the 1$\sigma$ dispersion around the mean),
assuming an age of 13.5~Gyr for all the stars. These objects span the [Fe/H] range between $\sim -$1.3 and $\sim -$2.0~dex 
--corresponding to [M/H] 
between $\sim -$1.0 and $\sim -$1.7~dex for a typical [$\alpha$/Fe]$\sim$0.4, 
\citep[see e.g.][and references therein]{alp}-- 
a regime where theory predicts a very small variation in ${\rm [C/N]_{FDU}}$ with metallicity, even at old ages (see Fig.~\ref{calibration}) 
hence it is justified to consider a mean ${\rm [C/N]_{FDU}}$ for the sample.
We then considered three stars in NGC6791 
\citep{n6791}, M67 \citep{m67}, NGC6819 \citep[][excluding the Li-rich star that appears on the lower RGB of the cluster]{carlb}, 
and NGC752 \citep{n752} respectively, plus one star in NGC2506 \citep{n2506}. 
In case of multiple objects from the same cluster we took the mean value of ${\rm [C/N]_{FDU}}$ and the dispersion 
around the mean.
The evolutionary status of the observed open cluster stars (post FDU but pre-RGB bump) has been determined 
making use of the appropriate isochrones from \citet{basti}.

For NGC6791 we employed the distance modulus and age from \citet{n6791age}, and [Fe/H]=0.37 \citep{n6791}. 
In case of M67 we considered [Fe/H]=0.05, distance modulus and age all taken from \citet{m67age}, whilst for NGC6819 
we adopted [Fe/H]=0.09, distance and age from \citet{ifmr}, and 
for NGC752 we employed [Fe/H]=$-$0.02 \citep{n752}, t=1.7$\pm$0 Gyr, $(m-M)_V$=8.36 \citep[see discussion in][]{n752}.
In case of NGC2506 we assumed [Fe/H]=$-$0.24 \citep{n2506}, age and distance modulus from \citet{n2506age}.

For M67, NGC6819 and NGC2506, 
where the original papers either determined only the number ratio (C/N) 
(the case of M67 and NGC6819) or assumed (not determined with the same methods) 
a value for ${\rm (C/N)_{\odot}}$ from 
independent estimates of the solar (C/N) ratio (the case of NGC2506), 
we employed ${\rm \log(C/N)_{\odot}}$=0.58 from \citet{gn93}, as in the BaSTI calculations.
This causes an increase in the measured [C/N] by just 0.02~dex for NGC2506, compared to the value provided 
by \citet{n2506}. 
Both \citet{a09} and \citet{gs98} solar mixtures give ${\rm \log(C/N)_{\odot}}$=0.60, whilst 
\citet{caffau} estimate gives ${\rm \log(C/N)_{\odot}}$=0.64.

Figure~\ref{calibration} summarizes the comparison between the observational data described above and our reference
theoretical predictions. As anticipated before, the data cannot provide a stringent test of the theory, but overall
they confirm the predicted decrease in ${\rm [C/N]_{FDU}}$ with age.

\section{Comparison with other stellar models}

We compared our reference ${\rm [C/N]_{FDU}}$-[M/H]-t calibration with results from the 
PARSEC \citep{parsec} and STAREVOL \citep{lagarde} databases of stellar evolution
calculations, which provide also surface chemical abundances of several elements during the evolution 
of models of different mass and initial chemical composition.
The solar heavy element distribution differs amongst BASTI, PARSEC and STAREVOL. 
BaSTI calculations employ the \citet{gn93} distribution, whilst PARSEC and STAREVOL models adopt the
\citet{caffau} and \citet{a09} determinations, respectively. 
These three different estimates provide different values of the solar metal mass fraction, that we accounted for
in the calculations of [M/H], whilst the solar C/N ratio --which is the C/N ratio of the models' 
initial chemical composition and is also needed to calculate ${\rm [C/N]_{FDU}}$-- 
does not change by more than 0.06~dex, as discussed in the previous section.
PARSEC and STAREVOL calculations also include convective core overshooting during the MS as our reference calibration, 
although with different efficiencies. In addition, PARSEC models account for overshooting from the convective envelope and 
atomic diffusion, partially inhibited from the surface convective layers, as discussed in \citet{parsec}.

\begin{figure}
\centering
\includegraphics[scale=.4500]{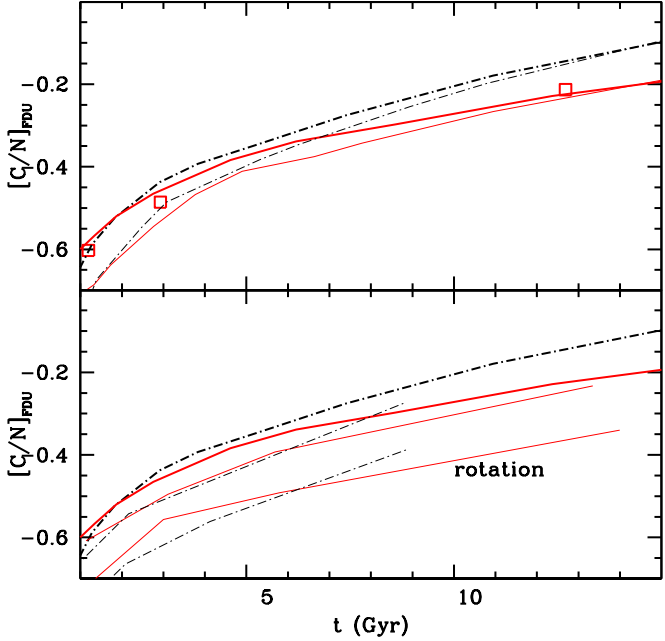}
\caption{Comparison between the BaSTI (thick lines) and PARSEC (thin lines, upper panel), STAREVOL (thin lines lower panel) 
  predictions for the ${\rm [C/N]_{FDU}}$-t relation at selected metallicities.
  BaSTI results for [M/H]=0.06 (solid line) and [M/H]=$-$0.66 (dash-dotted line) are displayed together
  with PARSEC results for [M/H]=$-$0.02 (solid line) and [M/H]=$-$0.58 (dash-dotted line), and STAREVOL results for 
 [M/H]=0.0 (solid line) and [M/H]=$-$0.56 (dash-dotted line).  
 STAREVOL models with rotation for the same metallicities are also displayed (with the same line styles).
 In the upper panel squares denote BaSTI calculations for [M/H]=0.04 
  and the \citet{caffau} solar heavy element distribution (see text for details).}
\label{parlag}
\end{figure}

The top panel of Fig.~\ref{parlag} compares our ${\rm [C/N]_{FDU}}$-t relations for [M/H]=0.06 (Z=0.02, Y=0.273) and $-$0.66~dex (Z=0.004, Y=0.251), 
with PARSEC results for
[M/H]=$-$0.02 (Z=0.014, Y=0.273) and $-$0.58~dex (Z=0.004, y=0.256) -- the BaSTI and PARSEC models are not calculated 
for exactly the same [M/H] grid.
There is good quantitative agreement between the two sets of results for ages larger than $\sim9-10$~Gyr, that deteriorates
towards younger ages. In general PARSEC models predict lower ${\rm [C/N]_{FDU}}$ values at a given t and [M/H], 
typically by $\sim$0.05~dex at t$\sim$5~Gyr, increasing to $\sim$0.10~dex at t$\sim$1.0~Gyr.
However, what is more relevant is the corresponding age difference for a given ${\rm [C/N]_{FDU}}$, at fixed [M/H].
PARSEC ages are systematically higher, 
with differences of the order of $\sim1.5-2.0$~Gyr for t between $\sim$4 and $\sim$8~Gyr. At older ages the difference decreases 
sharply to almost zero,
whereas it is of the order of $\sim$0.5-1~Gyr at ages below $\sim$4~Gyr.

By comparing ${\rm [C/N]_{FDU}}$ and FDU ages for various masses at fixed [M/H], we found that the 
discrepancy is mainly driven by lower ${\rm [C/N]_{FDU}}$ values at fixed mass for the PARSEC models. 
As a test we then calculated some models with MS core overshooting 
(without diffusion nor envelope overshooting) as in our BaSTI reference calibration, but 
with the \citet{caffau} solar metal distribution, for $Z$=0.0172 and $Y$=0.270, which is the initial solar composition determined by  
calibrating the standard solar model (with diffusion included). This initial chemical composition corresponds to [M/H]=0.04, 
and the resulting ${\rm [C/N]_{FDU}}$ 
values as a function of age are also reported in Fig.~\ref{parlag}. Just the effect of the solar metal distribution does not seem to be able to 
explain the differences with PARSEC results at the similar metallicity [M/H]=$-$0.02.

PARSEC calculations include overshooting from the convective envelopes that increases 
when the stellar mass increases \citep[see the discussion in][]{parsec}, and we tested also this effect. 
By employing the same envelope overshooting prescription as in the PARSEC calculations  
--${\rm \lambda_{ov}=0.7\times{H_P}}$ for masses above 1.4${\rm M_{\odot}}$, decreasing at lower masses down to ${\rm \lambda_{ov}=0.05\times{H_P}}$ when 
M$<1.1{\rm M_{\odot}}$)-- we obtained a decrease in ${\rm [C/N]_{FDU}}$ at fixed age 
of the order of 0.03~dex for ages below $\sim$3Gyr, this variation getting smaller at higher ages. 
Adding the 0.03~dex decrease in ${\rm [C/N]_{FDU}}$ to the square symbols in Fig.~\ref{parlag} at the ages $\sim$1 and $\sim$3~Gyr moves the points  
closer to the PARSEC results, still without matching them. 
In conclusion, part of the disagreement between the two calibrations seem to be due to the different metal distribution 
and the effect of envelope overshooting, even though some additional structural differences in the models --maybe related to the 
different efficiency of convective core overshooting during the MS-- play a role.

The lower panel of Fig.~\ref{parlag} displays a comparison with 
the STAREVOL calculations for [M/H]=0.0 (Z=0.014, Y=0.266) and [M/H]=$-$0.56 (Z=0.004, Y=0.253), 
respectively,
without and with the inclusion of rotation. Both sets of calculations include also thermoaline mixing, that sets in after the RGB bump, and is 
irrelevant for our comparisons.
The shape of the ${\rm [C/N]_{FDU}}$-t relation is less smooth than the case of BaSTI models, due to a coarser grid of masses.
STAREVOL calculations without rotation predict lower ${\rm [C/N]_{FDU}}$ at fixed age and [M/H], typically by $\sim$0.05~dex.  
The main culprit for these differences is the longer lifetime of STAREVOL models of a given mass and [M/H].
For a given ${\rm [C/N]_{FDU}}$, at fixed [M/H].
STAREVOL calibration provides therefore ages systematically higher than our calibration, 
with differences of the order of $\sim$1.5~Gyr, decreasing to $\sim0.5$ for ages below $\sim$2~Gyr. 

The STAREVOL models with rotation have been calculated with an initial rotational velocity equal to 45\% of the critical velocity for each mass,
and display much lower ${\rm [C/N]_{FDU}}$ values at a given t and [M/H], with differences up to $\sim$0.20~dex compared to our reference calibration.
This is due mainly to rotational mixing that pushes N from the inner regions
at CNO equilibrium (higher N abundance
at CNO equilibrium compared to the NO equilibrium) into overlying layers later to 
be engulfed by convection during the FDU. At the same time C is pushed inwards from the envelope, given
that CN and CNO equilibrium abundances are lower than the initial ones
\citep[see, e.g., Fig.11 from][]{lagarde}. 
The effect of rotation on lifetimes at the FDU is very small. In this mass range rotation causes a small increase in the 
MS lifetime, together with a shortening of the RGB timescales \citep[see discussion in][]{lagarde}, compared to non-rotating models.

A comparison of Figs.~\ref{parlag} and ~\ref{calibration} shows that at the lower age end of the ${\rm [C/N]_{FDU}}$-t calibration 
the existing sparse observational data cannot strongly confirm or disprove any of these alternative calibrations. However,  
at old ages models with rotation seem to largely underestimate ${\rm [C/N]_{FDU}}$ compared to NGC6791, if we assume 
that the scaling of ${\rm [C/N]_{FDU}}$ with [M/H] is similar to BaSTI results between solar 
and twice solar metallicity (almost negligible difference), given that [M/H]=0 is 
the upper boundary of STAREVOL metallicity range.

\section{Summary}
 
We have analysed the use of ${\rm [C/N]_{FDU}}$ as age indicator 
for RGB stars, and derived a theoretical calibration of ${\rm [C/N]_{FDU}}$ in terms of total metallicity and age,
for ages larger than ${\rm t\sim}$1~Gyr, based on the BaSTI stellar models with convective core overshooting.
In terms of general qualitative properties, lower metallicities imply higher ${\rm [C/N]_{FDU}}$ at old ages, but  
${\rm [C/N]_{FDU}}$ becomes increasingly less affected by [M/H] when t decreases, being 
almost independent of metallicity below $\sim$3~Gyr.

We have tested the robustness of our calibration against variations in the initial heavy element distribution (scaled-solar 
vs $\alpha$-enhanced), efficiency of overshooting from MS convective cores and from the convective envelopes, variations 
in the initial He abundance and in the mixing length parameter ${\rm \alpha_{\rm ml}}$.
The small sample of available measurements of ${\rm [C/N]_{FDU}}$ at least qualitatively confirm the overall trend of 
the predicted ${\rm [C/N]_{FDU}}$ with age and [M/H]. 

For ages ranging between 1 and 14~Gyr, ${\rm [C/N]_{FDU}}$ changes  
by at most $\sim$0.60~dex at the lowest metallicities, and by just $\sim$0.35~dex at supersolar [M/H]. For typical 
uncertainties of 0.10-0.15~dex on the observed abundance ratios, 
${\rm [C/N]_{FDU}}$-[M/H]-t relationships cannot therefore give very precise ages of individual stars, but may  
provide a general chronology for the formation of composite populations, when large samples  
of RGB stars with ${\rm [C/N]_{FDU}}$ measurements are available, and add further constraints to analyses of RGB samples   
from the log($g$)-log(${T_{\mathrm eff}}$) diagrams.  

Comparisons with the ${\rm [C/N]_{FDU}}$-[M/H]-t relations from the PARSEC stellar models 
display ${\rm [C/N]_{FDU}}$ differences at fixed t and [M/H] that increase with decreasing age, whilst 
STAREVOL models without rotation show a more constant offset of ${\rm [C/N]_{FDU}}$. 
Age differences up to $\sim$2~Gyr (PARSEC and STAREVOL ages being systematically higher) for a given   
${\rm [C/N]_{FDU}}$ and [M/H] pair are possible, mainly in the age range between $\sim$3-4 and $\sim$8~Gyr.
STAREVOL calculations with rotation display a further decrease in ${\rm [C/N]_{FDU}}$ at 
a given t and [M/H], that has an even larger impact on the ages derived from ${\rm [C/N]_{FDU}}$.

Additional accurate spectroscopic measurements of ${\rm [C/N]_{FDU}}$ in star clusters with well established 
ages and metallicities are necessary to better test the theoretical calibrations of this age indicator.

\begin{acknowledgements}

SC and AMP warmly thank financial support by PRIN-INAF2014 (PI: S. Cassisi), and by Economy and Competitiveness Ministry of the 
Kingdom of Spain (grant AYA2013-42781-P). AP acknowledges financial support by PRIN-INAF2013 (PI: L. Bedin).

\end{acknowledgements}

\bibliographystyle{aa}

\begin{thebibliography}{37}
\expandafter\ifx\csname natexlab\endcsname\relax\def\natexlab#1{#1}\fi

\bibitem[{{Angelou} {et~al.}(2015){Angelou}, {D'Orazi}, {Constantino},
  {Church}, {Stancliffe}, \& {Lattanzio}}]{a15}
{Angelou}, G.~C., {D'Orazi}, V., {Constantino}, T.~N., {et~al.} 2015, \mnras,
  450, 2423

\bibitem[{{Asplund} {et~al.}(2009){Asplund}, {Grevesse}, {Sauval}, \&
  {Scott}}]{a09}
{Asplund}, M., {Grevesse}, N., {Sauval}, A.~J., \& {Scott}, P. 2009, \araa, 47,
  481

\bibitem[{{Bedding} {et~al.}(2010){Bedding}, {Huber}, {Stello}, {Elsworth},
  {Hekker}, {Kallinger}, {Mathur}, {Mosser}, {Preston}, {Ballot}, {Barban},
  {Broomhall}, {Buzasi}, {Chaplin}, {Garc{\'{\i}}a}, {Gruberbauer}, {Hale}, {De
  Ridder}, {Frandsen}, {Borucki}, {Brown}, {Christensen-Dalsgaard},
  {Gilliland}, {Jenkins}, {Kjeldsen}, {Koch}, {Belkacem}, {Bildsten}, {Bruntt},
  {Campante}, {Deheuvels}, {Derekas}, {Dupret}, {Goupil}, {Hatzes}, {Houdek},
  {Ireland}, {Jiang}, {Karoff}, {Kiss}, {Lebreton}, {Miglio}, {Montalb{\'a}n},
  {Noels}, {Roxburgh}, {Sangaralingam}, {Stevens}, {Suran}, {Tarrant}, \&
  {Weiss}}]{keplera}
{Bedding}, T.~R., {Huber}, D., {Stello}, D., {et~al.} 2010, \apjl, 713, L176

\bibitem[{{Bellini} {et~al.}(2010){Bellini}, {Bedin}, {Piotto}, {Salaris},
  {Anderson}, {Brocato}, {Ragazzoni}, {Ortolani}, {Bonanos}, {Platais},
  {Gilliland}, {Raimondo}, {Bragaglia}, {Tosi}, {Gallozzi}, {Testa},
  {Kochanek}, {Giallongo}, {Baruffolo}, {Farinato}, {Diolaiti}, {Speziali},
  {Carraro}, \& {Yadav}}]{m67age}
{Bellini}, A., {Bedin}, L.~R., {Piotto}, G., {et~al.} 2010, \aap, 513, A50

\bibitem[{{B{\"o}cek Topcu} {et~al.}(2015){B{\"o}cek Topcu}, {Af{\c s}ar},
  {Schaeuble}, \& {Sneden}}]{n752}
{B{\"o}cek Topcu}, G., {Af{\c s}ar}, M., {Schaeuble}, M., \& {Sneden}, C. 2015,
  \mnras, 446, 3562

\bibitem[{{Bragaglia} {et~al.}(2014){Bragaglia}, {Sneden}, {Carretta},
  {Gratton}, {Lucatello}, {Bernath}, {Brooke}, \& {Ram}}]{n6791}
{Bragaglia}, A., {Sneden}, C., {Carretta}, E., {et~al.} 2014, \apj, 796, 68

\bibitem[{{Bressan} {et~al.}(2012){Bressan}, {Marigo}, {Girardi}, {Salasnich},
  {Dal Cero}, {Rubele}, \& {Nanni}}]{parsec}
{Bressan}, A., {Marigo}, P., {Girardi}, L., {et~al.} 2012, \mnras, 427, 127

\bibitem[{{Brogaard} {et~al.}(2012){Brogaard}, {VandenBerg}, {Bruntt},
  {Grundahl}, {Frandsen}, {Bedin}, {Milone}, {Dotter}, {Feiden}, {Stetson},
  {Sandquist}, {Miglio}, {Stello}, \& {Jessen-Hansen}}]{n6791age}
{Brogaard}, K., {VandenBerg}, D.~A., {Bruntt}, H., {et~al.} 2012, \aap, 543,
  A106

\bibitem[{{Caffau} {et~al.}(2011){Caffau}, {Ludwig}, {Steffen}, {Freytag}, \&
  {Bonifacio}}]{caffau}
{Caffau}, E., {Ludwig}, H.-G., {Steffen}, M., {Freytag}, B., \& {Bonifacio}, P.
  2011, \solphys, 268, 255

\bibitem[{{Carlberg} {et~al.}(2015){Carlberg}, {Smith}, {Cunha}, {Majewski},
  {M{\'e}sz{\'a}ros}, {Shetrone}, {Allende Prieto}, {Bizyaev}, {Stassun},
  {Fleming}, {Zasowski}, {Hearty}, {Nidever}, {Schneider}, {Holtzman}, \&
  {Frinchaboy}}]{carlb}
{Carlberg}, J.~K., {Smith}, V.~V., {Cunha}, K., {et~al.} 2015, \apj, 802, 7

\bibitem[{{Carretta} {et~al.}(2000){Carretta}, {Gratton}, \& {Sneden}}]{alp}
{Carretta}, E., {Gratton}, R.~G., \& {Sneden}, C. 2000, \aap, 356, 238

\bibitem[{{Cassisi} {et~al.}(2011){Cassisi}, {Mar{\'{\i}}n-Franch}, {Salaris},
  {Aparicio}, {Monelli}, \& {Pietrinferni}}]{cassisi:11}
{Cassisi}, S., {Mar{\'{\i}}n-Franch}, A., {Salaris}, M., {et~al.} 2011, \aap,
  527, A59

\bibitem[{{Cassisi} \& {Salaris}(2013)}]{cs13}
{Cassisi}, S. \& {Salaris}, M. 2013, {Old Stellar Populations: How to Study the
  Fossil Record of Galaxy Formation}

\bibitem[{{Chaplin} {et~al.}(2010){Chaplin}, {Appourchaux}, {Elsworth},
  {Garc{\'{\i}}a}, {Houdek}, {Karoff}, {Metcalfe}, {Molenda-{\.Z}akowicz},
  {Monteiro}, {Thompson}, {Brown}, {Christensen-Dalsgaard}, {Gilliland},
  {Kjeldsen}, {Borucki}, {Koch}, {Jenkins}, {Ballot}, {Basu}, {Bazot},
  {Bedding}, {Benomar}, {Bonanno}, {Brand{\~a}o}, {Bruntt}, {Campante},
  {Creevey}, {Di Mauro}, {Do{\v g}an}, {Dreizler}, {Eggenberger}, {Esch},
  {Fletcher}, {Frandsen}, {Gai}, {Gaulme}, {Handberg}, {Hekker}, {Howe},
  {Huber}, {Korzennik}, {Lebrun}, {Leccia}, {Martic}, {Mathur}, {Mosser},
  {New}, {Quirion}, {R{\'e}gulo}, {Roxburgh}, {Salabert}, {Schou}, {Sousa},
  {Stello}, {Verner}, {Arentoft}, {Barban}, {Belkacem}, {Benatti}, {Biazzo},
  {Boumier}, {Bradley}, {Broomhall}, {Buzasi}, {Claudi}, {Cunha}, {D'Antona},
  {Deheuvels}, {Derekas}, {Garc{\'{\i}}a Hern{\'a}ndez}, {Giampapa}, {Goupil},
  {Gruberbauer}, {Guzik}, {Hale}, {Ireland}, {Kiss}, {Kitiashvili},
  {Kolenberg}, {Korhonen}, {Kosovichev}, {Kupka}, {Lebreton}, {Leroy},
  {Ludwig}, {Mathis}, {Michel}, {Miglio}, {Montalb{\'a}n}, {Moya}, {Noels},
  {Noyes}, {Pall{\'e}}, {Piau}, {Preston}, {Roca Cort{\'e}s}, {Roth}, {Sato},
  {Schmitt}, {Serenelli}, {Silva Aguirre}, {Stevens}, {Su{\'a}rez}, {Suran},
  {Trampedach}, {Turck-Chi{\`e}ze}, {Uytterhoeven}, {Ventura}, \&
  {Wilson}}]{keplerb}
{Chaplin}, W.~J., {Appourchaux}, T., {Elsworth}, Y., {et~al.} 2010, \apjl, 713,
  L169

\bibitem[{{Gilmore} {et~al.}(2012){Gilmore}, {Randich}, {Asplund}, {Binney},
  {Bonifacio}, {Drew}, {Feltzing}, {Ferguson}, {Jeffries}, {Micela}, \&
  et~al.}]{gaiaeso}
{Gilmore}, G., {Randich}, S., {Asplund}, M., {et~al.} 2012, The Messenger, 147,
  25

\bibitem[{{Gilroy} \& {Brown}(1991)}]{m67}
{Gilroy}, K.~K. \& {Brown}, J.~A. 1991, \apj, 371, 578

\bibitem[{{Gratton} {et~al.}(2000){Gratton}, {Sneden}, {Carretta}, \&
  {Bragaglia}}]{g00}
{Gratton}, R.~G., {Sneden}, C., {Carretta}, E., \& {Bragaglia}, A. 2000, \aap,
  354, 169

\bibitem[{{Grevesse} \& {Noels}(1993)}]{gn93}
{Grevesse}, N. \& {Noels}, A. 1993, in Origin and Evolution of the Elements,
  ed. N.~{Prantzos}, E.~{Vangioni-Flam}, \& M.~{Casse}, 15--25

\bibitem[{{Grevesse} \& {Sauval}(1998)}]{gs98}
{Grevesse}, N. \& {Sauval}, A.~J. 1998, \ssr, 85, 161

\bibitem[{{Holtzman} {et~al.}(2015){Holtzman}, {Shetrone}, {Johnson}, {Allende
  Prieto}, {Anders}, {Andrews}, {Beers}, {Bizyaev}, {Blanton}, {Bovy},
  {Carrera}, {Cunha}, {Eisenstein}, {Feuillet}, {Frinchaboy}, {Galbraith-Frew},
  {Garcia Perez}, {Anibal Garcia Hernandez}, {Hasselquist}, {Hayden}, {Hearty},
  {Ivans}, {Majewski}, {Martell}, {Meszaros}, {Muna}, {Nidever}, {Nguyen},
  {O'Connell}, {Pan}, {Pinsonneault}, {Robin}, {Schiavon}, {Shane}, {Sobeck},
  {Smith}, {Troup}, {Weinberg}, {Wilson}, {Wood-Vasey}, {Zamora}, \&
  {Zasowski}}]{apogee}
{Holtzman}, J.~A., {Shetrone}, M., {Johnson}, J.~A., {et~al.} 2015, ArXiv
  e-prints

\bibitem[{{Karakas} \& {Lattanzio}(2014)}]{kl}
{Karakas}, A.~I. \& {Lattanzio}, J.~C. 2014, \pasa, 31, 30

\bibitem[{{Korn} {et~al.}(2006){Korn}, {Grundahl}, {Richard}, {Barklem},
  {Mashonkina}, {Collet}, {Piskunov}, \& {Gustafsson}}]{korn}
{Korn}, A.~J., {Grundahl}, F., {Richard}, O., {et~al.} 2006, \nat, 442, 657

\bibitem[{{Lagarde} {et~al.}(2012){Lagarde}, {Decressin}, {Charbonnel},
  {Eggenberger}, {Ekstr{\"o}m}, \& {Palacios}}]{lagarde}
{Lagarde}, N., {Decressin}, T., {Charbonnel}, C., {et~al.} 2012, \aap, 543,
  A108

\bibitem[{{Marconi} {et~al.}(1997){Marconi}, {Hamilton}, {Tosi}, \&
  {Bragaglia}}]{n2506age}
{Marconi}, G., {Hamilton}, D., {Tosi}, M., \& {Bragaglia}, A. 1997, \mnras,
  291, 763

\bibitem[{{Masseron} \& {Gilmore}(2015)}]{mt15}
{Masseron}, T. \& {Gilmore}, G. 2015, \mnras, 453, 1855

\bibitem[{{Michaud} {et~al.}(2007){Michaud}, {Richer}, \& {Richard}}]{mrr}
{Michaud}, G., {Richer}, J., \& {Richard}, O. 2007, \apj, 670, 1178

\bibitem[{{Mikolaitis} {et~al.}(2011){Mikolaitis}, {Tautvai{\v s}ien{\.e}},
  {Gratton}, {Bragaglia}, \& {Carretta}}]{n2506}
{Mikolaitis}, {\v S}., {Tautvai{\v s}ien{\.e}}, G., {Gratton}, R., {Bragaglia},
  A., \& {Carretta}, E. 2011, \mnras, 416, 1092

\bibitem[{{Mucciarelli} {et~al.}(2011){Mucciarelli}, {Salaris}, {Lovisi},
  {Ferraro}, {Lanzoni}, {Lucatello}, \& {Gratton}}]{msl11}
{Mucciarelli}, A., {Salaris}, M., {Lovisi}, L., {et~al.} 2011, \mnras, 412, 81

\bibitem[{{Pietrinferni} {et~al.}(2004){Pietrinferni}, {Cassisi}, {Salaris}, \&
  {Castelli}}]{basti}
{Pietrinferni}, A., {Cassisi}, S., {Salaris}, M., \& {Castelli}, F. 2004, \apj,
  612, 168

\bibitem[{{Pietrinferni} {et~al.}(2006){Pietrinferni}, {Cassisi}, {Salaris}, \&
  {Castelli}}]{bastialpha}
{Pietrinferni}, A., {Cassisi}, S., {Salaris}, M., \& {Castelli}, F. 2006, \apj,
  642, 797

\bibitem[{{Rodrigues} {et~al.}(2014){Rodrigues}, {Girardi}, {Miglio},
  {Bossini}, {Bovy}, {Epstein}, {Pinsonneault}, {Stello}, {Zasowski}, {Prieto},
  {Chaplin}, {Hekker}, {Johnson}, {M{\'e}sz{\'a}ros}, {Mosser}, {Anders},
  {Basu}, {Beers}, {Chiappini}, {da Costa}, {Elsworth}, {Garc{\'{\i}}a},
  {P{\'e}rez}, {Hearty}, {Maia}, {Majewski}, {Mathur}, {Montalb{\'a}n},
  {Nidever}, {Santiago}, {Schultheis}, {Serenelli}, \& {Shetrone}}]{r14}
{Rodrigues}, T.~S., {Girardi}, L., {Miglio}, A., {et~al.} 2014, \mnras, 445,
  2758

\bibitem[{{Salaris} {et~al.}(1993){Salaris}, {Chieffi}, \& {Straniero}}]{scs}
{Salaris}, M., {Chieffi}, A., \& {Straniero}, O. 1993, \apj, 414, 580

\bibitem[{{Salaris} {et~al.}(2009){Salaris}, {Serenelli}, {Weiss}, \& {Miller
  Bertolami}}]{ifmr}
{Salaris}, M., {Serenelli}, A., {Weiss}, A., \& {Miller Bertolami}, M. 2009,
  \apj, 692, 1013

\bibitem[{{Serenelli} {et~al.}(2013){Serenelli}, {Bergemann}, {Ruchti}, \&
  {Casagrande}}]{s13}
{Serenelli}, A.~M., {Bergemann}, M., {Ruchti}, G., \& {Casagrande}, L. 2013,
  \mnras, 429, 3645

\bibitem[{{Valle} {et~al.}(2015){Valle}, {Dell'Omodarme}, {Prada Moroni}, \&
  {Degl'Innocenti}}]{v15}
{Valle}, G., {Dell'Omodarme}, M., {Prada Moroni}, P.~G., \& {Degl'Innocenti},
  S. 2015, \aap, 575, A12

\bibitem[{{Yanny} {et~al.}(2009){Yanny}, {Rockosi}, {Newberg}, {Knapp},
  {Adelman-McCarthy}, {Alcorn}, {Allam}, {Allende Prieto}, {An}, {Anderson},
  {Anderson}, {Bailer-Jones}, {Bastian}, {Beers}, {Bell}, {Belokurov},
  {Bizyaev}, {Blythe}, {Bochanski}, {Boroski}, {Brinchmann}, {Brinkmann},
  {Brewington}, {Carey}, {Cudworth}, {Evans}, {Evans}, {Gates}, {G{\"a}nsicke},
  {Gillespie}, {Gilmore}, {Nebot Gomez-Moran}, {Grebel}, {Greenwell}, {Gunn},
  {Jordan}, {Jordan}, {Harding}, {Harris}, {Hendry}, {Holder}, {Ivans},
  {Ivezi{\v c}}, {Jester}, {Johnson}, {Kent}, {Kleinman}, {Kniazev},
  {Krzesinski}, {Kron}, {Kuropatkin}, {Lebedeva}, {Lee}, {French Leger},
  {L{\'e}pine}, {Levine}, {Lin}, {Long}, {Loomis}, {Lupton}, {Malanushenko},
  {Malanushenko}, {Margon}, {Martinez-Delgado}, {McGehee}, {Monet}, {Morrison},
  {Munn}, {Neilsen}, {Nitta}, {Norris}, {Oravetz}, {Owen}, {Padmanabhan},
  {Pan}, {Peterson}, {Pier}, {Platson}, {Re Fiorentin}, {Richards}, {Rix},
  {Schlegel}, {Schneider}, {Schreiber}, {Schwope}, {Sibley}, {Simmons},
  {Snedden}, {Allyn Smith}, {Stark}, {Stauffer}, {Steinmetz}, {Stoughton},
  {SubbaRao}, {Szalay}, {Szkody}, {Thakar}, {Sivarani}, {Tucker}, {Uomoto},
  {Vanden Berk}, {Vidrih}, {Wadadekar}, {Watters}, {Wilhelm}, {Wyse}, {Yarger},
  \& {Zucker}}]{segue}
{Yanny}, B., {Rockosi}, C., {Newberg}, H.~J., {et~al.} 2009, \aj, 137, 4377

\bibitem[{{Zhao} {et~al.}(2012){Zhao}, {Zhao}, {Chu}, {Jing}, \&
  {Deng}}]{lamost}
{Zhao}, G., {Zhao}, Y.-H., {Chu}, Y.-Q., {Jing}, Y.-P., \& {Deng}, L.-C. 2012,
  Research in Astronomy and Astrophysics, 12, 723

\end{thebibliography}

\end{document}